\documentclass[hyper]{JHEP} 

\usepackage{epsfig}

\newcommand\fverb{\setbox\pippobox=\hbox\bgroup\verb}
\newcommand\fverbdo{\egroup\medskip\noindent%
			\fbox{\unhbox\pippobox}\ }
\newcommand\fverbit{\egroup\item[\fbox{\unhbox\pippobox}]}
\newbox\pippobox
\title{A Toy Model of Closed String Tachyon
Effective Action}
\author{by J. Kluso\v{n}\\
	 Department of Theoretical Physics and Astrophysics\\
                   Faculty of Science, Masaryk University\\
Kotl\'{a}\v{r}sk\'{a} 2, 611 37, Brno\\
Czech Republic\\
	E-mail: \email{klu@physics.muni.cz}}
\preprint{\hepth{0404208}}

\abstract{In this paper we propose the toy model
of the closed string tachyon effective action
that has  marginal
tachyon profile as its exact solution in case of
constant or linear dilaton background. 
Then we will apply
 this  model for description of two dimensional bosonic string theory. 
We will find that the background configuration with the
spatial dependent linear dilaton, flat spacetime metric and
marginal tachyon profile is the exact solution of our model
even if we take into account 
 backreaction of tachyon on dilaton and on metric.}
\keywords{Closed string tachyon}

\def\bb{\mathbf{B}}

\def\ss{\sin \frac{\tau}{\sqrt{2}}}
\def\st{\sinh \frac{\tau}{\sqrt{2}}}
\def\st2{\sinh^2 \frac{\tau}{\sqrt{2}}}
\def\ss2{\sin^2 \frac{\tau}{\sqrt{2}}}

\begin{document}
\section{Introduction}\label{first}
The study of open string tachyon condensation
has led to the important insight into the non-perturbative
character of open string theory
\footnote{For  reviews of open string tachyon 
condensation, see \cite{Sen:1999mg,
Nakayama:2004vk,Taylor:2002uv,Arefeva:2001ps,
Ohmori:2001am,Schwarz:1999vu,Lerda:1999um,Taylor:2003gn}.}.
On the other hand the study of the 
closed string tachyon condensation
is more difficult \cite{Adams:2001sv,Dabholkar:2001if,Dabholkar:2001wn,
Harvey:2001wm,Martinec:2002tz,Gutperle:2002ki,
Armoni:2003va,He:2003yw,DaCunha:2003fm,Minwalla:2003hj,
Belopolsky:1994sk,Belopolsky:1994bj,Okawa:2004rh,
Dabholkar:2004ky}. The main problem is that spacetime
disappears altogether when the bulk tachyon condenses
and hence perturbation theory breaks down. 

One can hope that $c=1$ matrix model that provides
non-perturbative description of string theory
in $1+1$ dimensions 
(For review, see \cite{Kutasov:1991pv,Martinec:1991kn,
Klebanov:1991qa,Ginsparg:is,DiFrancesco:1993nw,
Jevicki:1993qn,Polchinski:1994mb}.) could be useful
laboratory for the study of the closed string tachyon
condensation. In fact, very interesting results considering time-dependent
tachyon condensation in two dimensions were obtained recently
in  
\cite{Karczmarek:2003pv,Karczmarek:2004ph,Das:2004hw}.
 
Another approach to the problem  the time-dependent closed string
tachyon condensation was given in 
\cite{DaCunha:2003fm,Strominger:2003fn,
Kluson:2003xn,Schomerus:2003vv}. 
This approach is based on the fact that there 
exists exact, time-dependent solution of
the tachyon field theories describing homogeneous
tachyon condensation. The worldsheet conformal
field theory that describes this process is
governed by the action (suppressing spatial
directions and setting $\alpha'=1$)
\begin{equation}\label{ac1}
S=\frac{1}{4\pi}\int d^2\sigma \left(
-(\partial X^0)^2+4\pi\mu e^{2\beta X^0}
\right) \ .
\end{equation}
which has negative norm boson and central charge
$c=1-6q^2 \ ,q=\beta-1/\beta$. The potential term
in (\ref{ac1}) can be interpreted as 
closed string tachyon field that grows exponentially
in time.

In very interesting paper  \cite{Kutasov:2003er}
the possibility of the effective field theory
description of such a process was discussed 
with analogy with the successful effective
field theory description of the open string
tachyon condensation
\cite{Sen:1999md,Garousi:2000tr,Bergshoeff:2000dq,
Kluson:2000iy,Minahan:2000tg,Sen:2002nu,
Sen:2002an,Sen:2002qa,Sen:2003tm,Smedback:2003ur,Fotopoulos:2003yt,
Sen:2003zf,Niarchos:2004rw,Lambert:2002hk,Lambert:2001fa,
Kluson:2003sr,Kluson:2004pj}.
According to this paper we should search for the
analogy of the exact solution in open string
theory on unstable D-brane $T=T_+e^{t}+T_-e^{-t}$ that
describes time-dependent process of the tachyon condensation.
It was proposed in \cite{Kutasov:2003er} that such
a closed string solution could be the world sheet interaction
\begin{equation}
\delta L=\lambda \cosh 2x^0 \ .
\end{equation}
The problem with this term is that it is expected 
not to be truly marginal and one can expect
large backreaction on the dilaton and metric in
the late times when we turn on $\lambda$. On
the other hand it was argued that the interaction
\begin{equation}\label{closmar}
\delta L=\lambda e^{2x^0}
\end{equation}
is exactly marginal and could be analogue of the
rolling tachyon profile in open string theory  even
if it is not completely clear what suppress
the backreaction to the metric and dilaton to
the stress tensor of tachyon.

As is well known from the study
of non-critical string theory  nonzero
vacuum expectation value of tachyon is especially
important  in case of the linear dilaton.
For that reason it seems to
be natural to study the closed string 
tachyon dynamics in case
of nontrivial dilaton background. 
Recently we have proposed an effective action 
for D-brane in the linear dilaton background 
 \cite{Kluson:2004qy}.
This effective action has the rolling tachyon
solution on unstable D-brane in the linear
dilaton background as its exact solution. 
Since the tachyon profile on unstable
D-brane  has similar form as  the tachyon 
exponential  
profile in the closed string case it is natural
to apply the formalism developed in
\cite{Kluson:2004qy} for the construction
of the closed string tachyon effective
action that will have (\ref{closmar}) as
its exact solution. In fact we should
be  more modest and say that we
are proposing a toy model for the closed
string tachyon effective action since
our approach is not based on the
first principles of the string theory as
for example calculation of the tachyon
effective action from the  partition
function on the two sphere.
In fact this is the same situation as in
the open string case where it was argued
\cite{Fotopoulos:2003yt}
that the tachyon effective action that has
the marginal tachyon profile as its exact solution
should be related 
by some complicated field redefinition
 to the tachyon effective
action derived from the disk partition function. 
We mean that the similar tachyon field
redefinition that maps our proposed action
to the tachyon effective action derived
from the sphere partition function 
could exist as well and for that reason we
hope that our toy model could be useful
for the study of the closed string
tachyon condensation. 

The structure of this paper is as follows.
In the next section 
(\ref{second}) we  propose field theory
effective action    for the tachyon
field with the mass square  $\mu^2$ that has
 profile (\ref{closmar})
 as its exact solution
of the equation of motion. We will calculate
the stress energy tensor and dilaton source.
Then in section (\ref{third})
 we will apply our proposal to the
case of two dimensional string theory where
the tachyon effective action considerably
simplifies as a consequence of the fact
that   the potential term in the tachyon
effective action  vanishes.  Since generally
nonzero tachyon field forms sources for the
dilaton and for the graviton  we will
study the problem of the backreaction of
the tachyon on the metric and dilaton.
After inclusion of the additional
term to the tachyon effective action that
could not be derived from the requirement that
the rolling tachyon is the solution
of the equation of motion in two dimensional
theory 
we obtain the  action 
for dilaton, graviton and tachyon that has
the linear dilaton background in two dimensions
with flat Minkowski metric and with 
exponential tachyon profile as its exact solution
even if we take into account the backreaction
of the tachyon on dilaton and on metric.
We also find another time dependent solution
that has the same property. On the other hand
we will show  that the tachyon effective action
poses some exact solutions that  generate
nonzero dilaton source and stress energy tensor
and hence cannot be considered as an exact
solution of this toy model of 
two dimensional effective  field theory. 
 In section (\ref{fourth})
we will study  fluctuations around the
tachyon marginal profile. We will
find an effective action for fluctuations
that has plane waves corresponding to
propagating 
massless modes as its exact solution.

Finally, in conclusion (\ref{fifth})
 we outline our
results and suggest further  problems that
deserve to by studied. 
\section{Proposal for 
closed string tachyon effective action in 
constant and in the linear dilaton background}\label{second}
In this section we propose the effective
action for  massive field $T$ that in case of
constant dilaton $\Phi_0$  has
the leading order 
condition of marginality 
\footnote{Our convention is 
$\eta_{\mu\nu}=(-1,1,\dots,D-1)$ where 
$D$ is the number of spacetime dimensions. We
also have  $i,j=1,\dots, D-1$, where $x^i$ label
spatial directions.}
\begin{equation}
\partial_{\mu}\left[\eta^{\mu\nu}\partial_{\nu}T
\right]=-\mu^2T \ 
\end{equation}
Using results given in 
\cite{Kutasov:2003er,Smedback:2003ur,Niarchos:2004rw,
Kluson:2004qy} we now presume that
 the tachyon effective  action has the form
\begin{equation}\label{actpro}
S=-\int d^Dx \mathcal{L} \ , 
\mathcal{L}=\frac{1}{(1+kT^2)}\sqrt{\bb} \ ,
\bb=1+\mu^2T^2+\eta^{\mu\nu}\partial_{\mu}T
\partial_{\nu}T \ ,  
\end{equation}
where the unknown constant $k$ will be
determined from the requirement that
 following tachyon
profile  
\begin{equation}\label{ans1}
T=T_+e^{\beta_{\mu}^+x^{\mu}}+
T_-e^{\beta_{\mu}^-x^{\mu}}  \ ,
\beta_{\mu}^{\pm}\eta^{\mu\nu}
\beta_{\nu}^{\pm}=-\mu^2 
\end{equation}
with  $\beta^{\pm}_i=0$ is an exact solution
of the equation of motion that arises
from (\ref{actpro})
\begin{eqnarray}\label{eq1}
-\frac{2kT\sqrt{\bb}}{(1+kT^2)^2}+
\frac{\mu^2T}{(1+kT^2)\sqrt{\bb}}-\nonumber \\
\frac{
\partial_{\mu}\eta^{\mu\nu}\partial_{\nu}T}
{(1+kT^2)\sqrt{\bb}}+
\frac{2kT\partial_{\mu}T\eta^{\mu\nu}\partial_{\nu}T}
{(1+kT^2)^2\sqrt{\bb}}=0 \ ,
\end{eqnarray}
where we have presumed that  $\bb$ is constant.  This
assumption  holds for
the ansatz  (\ref{ans1}) since in this case
we have 
\begin{equation}
\bb=1+2T_+T_-e^{(\beta_++\beta_-)t}
(\mu^2+2(\beta_++\beta_-))= 
1+2T_+T_-\mu^2=const \ .
\end{equation}
Consequently the  equation of motion
(\ref{eq1}) reduces for (\ref{ans1}) to 
\begin{equation}
\frac{-2kT\sqrt{\bb}}{(1+kT^2)^2}+\frac{2\mu^2T}
{(1+kT^2)\sqrt{\bb}}+\frac{2kT(\bb-1-\mu^2T^2)}{(1+kT^2)^2
\sqrt{\bb}}=0 \ .
\end{equation}
The expression given above suggests that
we should take  $k=\mu^2$ in order to
obey equation of motion 
since then
\begin{equation}
\frac{2\mu^2T}{(1+\mu^2T^2)^2\sqrt{\bb}}
(-\bb+\bb-1-\mu^2T^2)+\frac{2\mu^2T}
{(1+\mu^2T^2)\sqrt{\bb}}=0 \ .
\end{equation}
 In summary, 
 the tachyon effective Lagrangian 
that has  tachyon profile 
\begin{equation}
T=\lambda e^{\beta_{\mu}x^{\mu}} \ ,
-\beta_{\mu}\eta^{\mu\nu}\beta_{\nu}=\mu^2
\end{equation}
as its exact solution takes the form
\begin{equation}\label{act3}
\mathcal{L}=\frac{A}{1+\mu^2T^2}\sqrt{\bb} \ ,
\end{equation}
where we have included possible numerical factor $A$
that cannot be determined from this analysis. 

Now we apply this strategy to the case
of the closed string tachyon in the linear dilaton
background. In this case the leading order
 condition of marginality
is 
\begin{equation}\label{conmar}
\partial_{\mu}[e^{-2\Phi}\eta^{\mu\nu}\partial_{\nu}T]
=-\mu^2T \ .
\end{equation}
Following arguments given in
\cite{Kluson:2004qy} we suggest 
that the  Lagrangian that has
the  tachyon profile obeying
(\ref{conmar}) as its exact solution
has the  form
\begin{equation}\label{actpro2}
\mathcal{L}=\frac{A}{1+kT^2e^{-2\Phi}}
\sqrt{\bb} \ ,
\bb=1+e^{-2\Phi}(T^2\mu^2+\eta^{\mu\nu}\partial_{\mu}T
\partial_{\nu}T-2T
\eta^{\mu\nu}\partial_{\mu}T\partial_{\nu}
\Phi) 
\end{equation}
so that the equation of motion is
\begin{eqnarray}\label{dileq}
-\frac{2kTe^{-2\Phi}\sqrt{\bb}}
{(1+kT^2e^{-2\Phi})^2}
+\frac{\mu^2Te^{-2\Phi}}{(1+kT^2e^{-2\Phi})\sqrt{\bb}}
-\partial_{\mu}\left[\frac{e^{-2\Phi}
\eta^{\mu\nu}\partial_{\nu}T}{
(1+kT^2e^{-2\Phi})\sqrt{\bb}}\right]-\nonumber\\
-\frac{e^{-2\Phi}\eta^{\mu\nu}\partial_{\mu}T
\partial_{\nu}\Phi}{(1+kT^2e^{-2\Phi})\sqrt{\bb}}
+\partial_{\mu}\left[\frac{Te^{-2\Phi}
\eta^{\mu\nu}\partial_{\nu}\Phi}{(1+kT^2e^{-2\Phi})
\sqrt{\bb}}\right]=0 \ . \nonumber \\
\end{eqnarray}
As in case of constant dilaton reviewed
above  we determine the
constant $k$ from the requirement that
the  tachyon  profile 
that obeys (\ref{conmar}) and for which
$\bb$ is constant 
\footnote{Generally not all tachyon
marginal profiles obeying (\ref{conmar}) imply
$\bb=const$.  On the other hand
we will see that for exponential tachyon profile
that obeys (\ref{conmar}) the expression $\bb$ is equal
to one. 
 For more detailed
discussion of this issue, see
 \cite{Kluson:2004qy}.}
is an exact solution of the equation of motion.
Using
the fact that  $\sqrt{\bb}=const$ the equation
of motion (\ref{dileq}) after some calculation
reduces to
\begin{eqnarray}
\frac{2kTe^{-2\Phi}}{(1+kT^2e^{-2\Phi})^2\sqrt{\bb}}
(-1-e^{-2\Phi}T^2(\mu^2-V_{\mu}V^{\mu}))
+\frac{2T e^{-2\Phi}}{(1+kT^2e^{-2\Phi})\sqrt{\bb}}
(\mu^2-V_{\mu}V^{\mu})=0 \ . 
\nonumber \\
\end{eqnarray}
We see that the previous equation has solution if 
we take 
$k=\mu^2-V_{\mu}V^{\mu}$ since then 
\begin{equation}
\frac{2(\mu^2-V_{\mu}V^{\mu})
Te^{-2\Phi}}{(1+kT^2e^{-2\Phi})^2\sqrt{\bb}}
(-1-e^{-2\Phi}T^2k))
+\frac{2T e^{-2\Phi}}{(1+kT^2e^{-2\Phi})\sqrt{\bb}}
(\mu^2-V_{\mu}V^{\mu})=0 \ .
\end{equation}
Consequently  our proposal for 
the closed string tachyon 
 effective action in bosonic string
theory ($\mu^2=4$)  in the
 linear
dilaton background  has the form
\begin{equation}\label{act2}
S=-\int d^Dx\mathcal{L} \ , 
\mathcal{L}=\frac{A}{
(1+e^{-2\Phi}T^2(4-V_{\mu}V^{\mu}))} 
\sqrt{\bb} \ . 
\end{equation}
We see that for $D=2$ the
 potential term in (\ref{act2})
vanishes
thanks to the
relation between dimension of spacetime
and the norm of the dilaton vector $V:$ 
$4-V_{\mu}V^{\mu}=4-(26-D)/6=\frac{D-2}{6}$
that holds in bosonic string theory. 
This fact will play significant role in the
next section. 

In order to calculate
the stress energy tensor and dilaton source
 we will proceed as follows. Firstly 
 we replace $V_{\mu}V^{\mu}$ with
$\partial_{\mu}\Phi\eta^{\mu\nu}\partial_{\nu}
\Phi$. Now we    
presume that the action (\ref{act2}) is
also valid for any  dilaton field. In the
same way we replace everywhere the flat spacetime
metric $\eta^{\mu\nu}$ with $g^{\mu\nu}$ and
also  $d^Dx$ with $d^Dx\sqrt{-g}$ and presume
that (\ref{act2}) is valid for general
metric $g$.  Then
 the stress energy tensor is equal to
\begin{equation}
T_{\mu\nu}=-g_{\mu\nu}\mathcal{L}+
2\frac{\delta \mathcal{L}}{\delta g^{\mu\nu}} \ .
\end{equation}
Using the Lagrangian 
 (\ref{act2}) we get 
\begin{eqnarray}\label{Tstress}
T_{\mu\nu}=-g_{\mu\nu}
\frac{A}{
(1+e^{-2\Phi}T^2(4-
\partial_{\mu}\Phi g^{\mu\nu}\partial_{\nu}
\Phi))}
\sqrt{\bb}
+\nonumber \\
+\frac{2Ae^{-2\Phi}T^2
\partial_{\mu}\Phi\partial_{\nu}\Phi}{
(1+e^{-2\Phi}T^2(4-
\partial_{\mu}\Phi g^{\mu\nu}\partial_{\nu}
\Phi))^2}
\sqrt{\bb}+
\nonumber \\
+\frac{Ae^{-2\Phi}(\partial_{\mu}T\partial_{\nu}T
-T\left(
\partial_{\mu}T\partial_{\nu}\Phi
+\partial_{\mu}\Phi\partial_{\nu}T
\right))}{
(1+e^{-2\Phi}T^2(4-
\partial_{\mu}\Phi g^{\mu\nu}\partial_{\nu}
\Phi))}
\sqrt{\bb} \nonumber \\
\end{eqnarray}
and the  the dilaton source 
$J_{\Phi}=\frac{\delta S}{\delta 
\Phi}$ 
\begin{eqnarray}\label{dilsor}
J_{\Phi}=-\frac{2A\sqrt{-g}e^{-2\Phi}T^2
(4-\partial_{\mu}\Phi g^{\mu\nu}
\partial_{\nu}\Phi)\sqrt{\bb}}
{(1+e^{-2\Phi}T^2(4-
\partial_{\mu}\Phi g^{\mu\nu}\partial_{\nu}
\Phi))^2}+
\nonumber \\
+\partial_{\mu}\left[
\frac{2A\sqrt{-g}e^{-2\Phi}T^2g^{\mu\nu}\partial_{\nu}\Phi
\sqrt{\bb}}
{(1+e^{-2\Phi}T^2(4-
\partial_{\mu}\Phi g^{\mu\nu}\partial_{\nu}
\Phi))^2}\right]+
\nonumber \\
+\frac{A\sqrt{-g}e^{-2\Phi}(4T^2+g^{\mu\nu}\partial_{\mu}T
\partial_{\nu}T-2Tg^{\mu\nu}\partial_{\mu}T
\partial_{\nu}\Phi)}{
(1+e^{-2\Phi}T^2(4-
\partial_{\mu}\Phi g^{\mu\nu}\partial_{\nu}
\Phi))\sqrt{\bb}}-
\nonumber \\
-\partial_{\mu}\left[\frac{A\sqrt{-g}e^{-2\Phi}
Tg^{\mu\nu}\partial_{\nu}T}
{(1+e^{-2\Phi}T^2(4-
\partial_{\mu}\Phi g^{\mu\nu}\partial_{\nu}
\Phi))\sqrt{\bb}}\right] \ . 
\nonumber \\
\end{eqnarray}
These expressions will be useful in the next
section when we use the action
(\ref{act2}) for the effective field
theory description of the tachyon
dynamics in two dimensions.   
\section{Toy model of two dimensional 
string theory}\label{third}
In this section we will formulate 
simple toy model  of two dimensional bosonic
 string theory. 
The starting point of this  model
is  the  generalisation of the tachyon
effective action  (\ref{act2}) to the
case of general metric $g_{\mu\nu}$ and
dilaton $\Phi$ as was performed in
the end of the previous section 
\begin{eqnarray}\label{Ttwo}
S_T=-A\int d^2x\frac{\sqrt{-g}}{
(1+e^{-2\Phi}T^2(4-g^{\mu\nu}\partial_{\mu}\Phi
\partial_{\nu}\Phi))}
\times \nonumber \\
\times \sqrt{1+e^{-2\Phi}\left(4T^2+g^{\mu\nu}
\partial_{\mu}T\partial_{\nu}T-2T
g^{\mu\nu}\partial_{\mu}T\partial_{\nu}\Phi
\right)} \ . \nonumber \\
\end{eqnarray}
The action for  metric and the
dilaton field is
\begin{equation}\label{metdil}
S_{g,\Phi}=-\int d^2x\sqrt{-g}e^{-2\Phi}
\left(16+R+4g^{\mu\nu}\partial_{\mu}\Phi
\partial_{\nu}\Phi\right) \ .
\end{equation}
The variation of the action $S=S_{g,\Phi}+S_T$
with respect to $g^{\mu\nu}$ gives
the equation of motion for $g_{\mu\nu}$
\begin{equation}\label{gequation}
e^{-2\Phi}\left(G_{\mu\nu}-
2g_{\mu\nu}\nabla^2\Phi+2\nabla_{\mu}
\nabla_{\nu}\Phi+2g_{\mu\nu}
\left(\nabla \Phi\right)^2-8g_{\mu\nu}\right)=
T_{\mu\nu}^T \ 
\end{equation}
and the variation with respect to $\Phi$ gives
\begin{equation}\label{phiequat}
\sqrt{-g}e^{-2\Phi}
\left[32+2R-8g^{\mu\nu}\partial_{\mu}\Phi
\partial_{\nu}\Phi\right]
+16e^{-2\Phi}\partial_{\mu}
\left[\sqrt{-g}g^{\mu\nu}\partial_{\nu}
\Phi\right]=J_{\Phi} \ ,
\end{equation}
where
\begin{equation}
T^T_{\mu\nu}=\frac{\delta S_T}{\delta g^{\mu\nu}} \ ,
J_{\Phi}=\frac{\delta S_T}{\delta \Phi} \ .
\end{equation}
The explicit form of the components of the
stress energy tensor and dilaton source is given in
(\ref{Tstress}) and in (\ref{dilsor}).

Let us consider standard linear dilaton
background configuration
\begin{equation}\label{linb}
\Phi=Vx \ , g_{\mu\nu}=\eta_{\mu\nu} \ ,
\end{equation}
where $x^0=t \ , x^1=x$.
Without inclusion of the backreaction of
$T$ the equation of motion for dilaton and metric
implies that $V^2=4$. On the other hand
we know that the exact solution of the 
equation of motion that arises from
(\ref{Ttwo}) is for the ansatz
(\ref{linb})   equal to
\begin{equation}\label{Tdil}
T=\mu e^{\beta x} \ , 
-\beta^2+2\beta V=4 \Rightarrow \beta=2 \ .
\end{equation}
Now we must ask the question whether
there is any backreaction of the tachyon
on the dilaton and metric. First of all,
for the ansatz (\ref{linb}) and (\ref{Tdil})
the dilaton current  $J_{\Phi}$
is equal to
\begin{eqnarray}
J_{\Phi}=\left[6\beta V-4V^2-2\beta^2\right]e^{-2\Phi}T^2=0 \ .
\nonumber \\
\end{eqnarray}
 This result implies that
the  nonzero tachyon condensate does not
modify the dilaton equation of motion and
hence ansatz (\ref{linb}), (\ref{Tdil})
 is  solution of
the equation of motion for dilaton and 
tachyon. As a last step we must show that
this ansatz also solves the equation of motion
for metric. In order to do this
we will calculate the stress energy
tensor   evaluated on the ansatz
(\ref{linb}), (\ref{Tdil}) and we 
find 
\begin{equation}
T_{00}^T=A \ ,
T_{01}=0 \ ,
T_{11}=A(-1+4\mu^2) \ . 
\end{equation}
However we would like to 
have such a toy model of two dimensional
string theory that has the background
corresponding to the conformal 
 Liouville field theory  (\ref{linb}), (\ref{Tdil}) 
as its exact solution. 
In order  find such an action we suggest to 
introduce additional  term into
the tachyon effective action that does
not have  impact on  the equation of motion for
the dilaton and tachyon as far as
two dimensional theory is 
considered.  

 More precisely,
let us consider following term
\begin{equation}\label{slambda}
S_{\Lambda}=-\int d^2x \sqrt{-g}\frac{\Lambda
}
{1+kT^2e^{-2\Phi}(4-\partial_{\mu}\Phi g^{\mu\nu}
\partial_{\nu}\Phi)} \ ,
\end{equation} 
where unknown constants $\Lambda \ , k$ should
be determined from the condition that total
stress energy tensor vanishes.  
As the first step we must confirm that
 the variation of this
term with respect to $T$ and $\Phi$
vanishes for 
the ansatz (\ref{linb}), (\ref{Tdil}).
 The variation
with  respect to $T$ gives
\begin{equation}\label{varTslambda}
\frac{\delta S_{\Lambda}}
{\delta T}= 2 
\int d^2x \sqrt{-g}\frac{\Lambda
e^{-2\Phi}kT(4-\partial_{\mu}\Phi g^{\mu\nu}
\partial_{\nu}\Phi)}
{(1+kT^2(4-\partial_{\mu}\Phi g^{\mu\nu}
\partial_{\nu}\Phi))^2}=0  \ ,
\end{equation} 
where we have used  $V_{\mu}V^{\mu}=4$. In the same way
one can show that
\begin{eqnarray}
\frac{\delta S_{\Lambda}}
{\delta \Phi}=
-2 
\int d^2x \sqrt{-g}\frac{\Lambda
ke^{-2\Phi}(4-\partial_{\mu}\Phi g^{\mu\nu}
\partial_{\nu}\Phi)}
{(1+kT^2(4-\partial_{\mu}\Phi g^{\mu\nu}
\partial_{\nu}\Phi))^2}+\nonumber \\
+
2\int d^2x \partial_{\mu}
\left[\frac{\sqrt{-g}k\Lambda T^2e^{-2\Phi}
g^{\mu\nu}\partial_{\nu}\Phi}
{(1+kT^2(4-\partial_{\mu}\Phi g^{\mu\nu}
\partial_{\nu}\Phi))^2}\right]
=0 \ , \nonumber \\
\end{eqnarray}
where the first term vanishes for 
$V_{\mu}V^{\mu}=4$ 
 and the second term is
equal to zero since the expression
in the bracket is constant. 
Finally let us calculate the variation
of (\ref{slambda}) with
respect to $g^{\mu\nu}$. Since the action has
the form $S=-\int d^2x \sqrt{-g}\mathcal{L}_{\Lambda}$
we get the components of the stress energy tensor as
\begin{eqnarray}
T_{\mu\nu}^{\Lambda}
=-g_{\mu\nu}\mathcal{L}_{\Lambda}+
2\frac{\delta \mathcal{L}_{\Lambda}}
{\delta g^{\mu\nu}}=\nonumber \\
=-g_{\mu\nu}\frac{\Lambda}
{1+kT^2e^{-2\Phi}(4-g^{\mu\nu}
\partial_{\mu}\Phi\partial_{\nu}\Phi)}
+2\frac{e^{-2\Phi}kT^2\Lambda \partial_{\mu}
\Phi\partial_{\nu}\Phi}{
(1+T^2e^{-2\Phi}k(4-g^{\mu\nu}
\partial_{\mu}\Phi\partial_{\nu}\Phi))^2}
\nonumber \\
\end{eqnarray}
that for (\ref{linb}) and (\ref{Tdil})
are equal to
\begin{equation}
T_{00}=\Lambda \ ,
T_{11}=-\Lambda+2k\Lambda\mu^2V^2=
-\Lambda+8k\Lambda \mu^2 \ .
\end{equation}
We see that for $\Lambda=-A \ ,k=\frac{1}{2}$
these terms precisely cancel the contribution
from the tachyon stress energy tensor. 

To conclude, we have found 
 two dimensional effective  action $S=S_T+S_{\Lambda}+
S_{\Phi,g}$ that has the background
(\ref{linb}) , (\ref{Tdil}) as its exact solution
even if we take into account the backreaction of
the tachyon on metric and dilaton.

Next question is whether our toy
model has another exact solution.  For example
we could hope to find 
an analogue of the  matrix cosmology solutions
\cite{Karczmarek:2003pv,Karczmarek:2004ph,Das:2004hw}. 
To begin with  let us consider 
following ansatz for the tachyon, dilaton 
and metric
\begin{eqnarray}\label{ansg}
T=\lambda e^{\beta_{\mu}x^{\mu}} \ ,
-\beta_{\mu}\eta^{\mu\nu}\beta_{\nu}+
2\beta_{\mu}\eta^{\mu\nu}V_{\nu}=4
\nonumber \\
\Phi=V_{\mu}x^{\mu} \ , 
V_{\mu}V^{\mu}=4 \ , g^{\mu\nu}=\eta^{\mu\nu} \ .
\nonumber \\
\end{eqnarray}
Since $T$ obeys (\ref{conmar}) and also 
for this ansatz $\bb=1$ it follows that
$T$ solves the equation of motion that
arises from (\ref{Ttwo}) and (\ref{slambda})
if we take also account
the  values of 
$g$ and $\Phi$ given in (\ref{ansg}). 
One  can also show
that the dilaton source is zero for (\ref{ansg})
\begin{eqnarray}
J_{\Phi}=2A\partial_{\mu}[e^{-2\Phi}T^2\eta^{\mu\nu}
V_{\nu}]-A\partial_{\mu}[e^{-2\Phi}
T^2\beta_{\nu}\eta^{\mu\nu}]
-A\partial_{\mu}\left[e^{-2\Phi}T^2\eta^{\mu\nu}
V_{\nu}\right]=\nonumber \\
=
2Ae^{-2\Phi}T^2[-\beta_{\mu}\beta^{\mu}+2\beta_{\mu}V^{\mu}
-V_{\mu}V^{\mu}]=
2Ae^{-2\Phi}T^2[4-V_{\mu}V^{\mu}]=0 \ 
\nonumber \\
\end{eqnarray}
and consequently the linear dilaton background
is not affected by the  nonzero condensate 
 of the tachyon. 

 On the other hand  we can expect  that
not all  values of $V, \beta$ are allowed
when we demand  the vanishing
of the stress energy tensor $T_{\mu\nu}=
T_{\mu\nu}^T+T_{\mu\nu}^{\Lambda}$. To see
this  note that diagonal components of
$T_{\mu\nu}$ 
are equal to 
\begin{eqnarray}
T_{00}=1+e^{-2\Phi}T^2(
2V_0^2+\beta_0^2-2\beta_0V_0)-1
-V_0^2e^{-2\Phi}T^2=
e^{-2\Phi}T^2(\beta_0^2-2\beta_0V_0
+V_0^2) \ , \nonumber \\
T_{11}=-1+e^{-2\Phi}T^2
(2V_1^2+\beta_1^2-2\beta_1V_1)+1
-V_1^2e^{-2\Phi}T^2=
(V_1^2+\beta_1^2-2\beta_1V_1)
e^{-2\Phi}T^2 \ . 
\nonumber \\
\end{eqnarray}
We see that the diagonal components of
the stress energy tensor vanish when 
\begin{equation}
\beta_{\mu}=V_{\mu} 
\end{equation}
that clearly obeys the condition 
for tachyon given in (\ref{ansg})
\begin{equation}
-\eta^{\mu\nu}\beta_{\mu}\beta_{\nu}+
2\eta^{\mu\nu}\beta_{\mu}V_{\nu}=
 -V_{\mu}V^{\mu}+2
V_{\mu}V^{\mu}=4 \ . 
\end{equation}
As a last check we confirm that the off-diagonal
components of the stress energy tensor  are
equal to zero too
\begin{eqnarray}
T_{01}=T_{10}=
e^{-2\Phi} T^2\left(2V_0V_1+\beta_0\beta_1
-(\beta_0V_1+V_0\beta_1)\right)-V_0V_1
e^{-2\Phi}T^2=0 \ . \nonumber \\
\end{eqnarray}
 In other
words the flat metric is solution of the equation
of motion.  We must however stress that
this exact solution is different from
the  solutions presented in 
\cite{Karczmarek:2003pv} since in this
paper the pure spatial dependent 
dilaton was considered and the tachyon
profile for $x\rightarrow -\infty$ 
has the form $T\sim xe^{2x}$. To see 
 whether such a tachyon profile is 
exact solution of the action (\ref{act2})
let us  consider an ansatz
\begin{equation}\label{T1}
T=(b+a_{\mu}x^{\mu})e^{\beta_{\mu}x^{\mu}} \ 
\end{equation}
together with flat metric and linear dilaton
background
\begin{equation}\label{T2}
g_{\mu\nu}=\eta_{\mu\nu} \ ,
\Phi=V_{\mu}x^{\mu} \ , V_{\mu}V^{\mu}=4 \  .
\end{equation}
Now we demand that the tachyon profile
(\ref{T1})
obeys 
(\ref{conmar}) which  implies
\begin{eqnarray}
\partial_{\mu}[e^{-2\Phi}\eta^{\mu\nu}
\partial_{\nu}T]=-4T \Rightarrow
\nonumber \\
T[\beta_{\mu}\beta^{\mu}-2V_{\mu}\beta^{\mu}
+4]+2e^{\beta_{\mu}x^{\mu}}
a_{\mu}(V^{\mu}-\beta^{\mu})=0
\Rightarrow
V_{\mu}=\beta_{\mu} \ 
\nonumber \\
\end{eqnarray}
and also 
\begin{eqnarray}
\bb=1+e^{-2\Phi}[(4T^2+\eta^{\mu\nu}
\partial_{\mu}T\partial_{\nu}T
-2e^{-2\Phi}TV_{\mu}\eta^{\mu\nu}\partial_{\nu}T
)+\nonumber \\
+2Ta_{\mu}\eta^{\mu\nu}[\beta_{\mu}-
V_{\mu}]e^{\beta x}+a_{\mu}a^{\mu}e^{2\beta x}]
=1+a_{\mu}a^{\mu} \ 
\nonumber \\
\end{eqnarray} 
using  the background configuration
(\ref{T2}). 
Then it
is clear that the tachyon profile
(\ref{T1}) is solution of
the equation of motion that
arises from the action (\ref{Ttwo}) 
if we also take into account (\ref{T2}).
On the other hand 
one can easily show that 
$J_{\Phi}$ is  nonzero for general $a_{\mu}$ and
hence generates source for the dilaton. 
For example, for spatial dependent 
dilaton and tachyon with $\Phi=2x \ ,T=(b+ax)e^{2x}$ 
one finds following dilaton source
\begin{equation}
J_{\Phi}=\frac{2A(ax+b)a[4a^2+2-
\sqrt{1+a^2}]}{\sqrt{1+a^2}}
-\frac{a^2A}{\sqrt{1+a^2}} \ 
\end{equation}
that diverges at asymptotic regions
$x\rightarrow \pm \infty$ and consequently
induces large backreaction on the dilaton.
In the same one can show that the
tachyon profile (\ref{T1}) induces nonzero components
of the stress energy tensor
 that implies that background (\ref{T2})
cannot be solution of the equation
of motion. In summary, the tachyon profile
(\ref{T1}) even if it is solution of the tachyon
equation of motion in flat spacetime and
in the linear dilaton background
  induces large backreaction
on dilaton and metric and hence its meaning
in the context of our model is unclear.
\section{Analysis of fluctuation}\label{fourth}
In this section we will study the fluctuations around
the classical solution in two dimensional string
theory. From the analysis performed in case
of the tachyon effective action for unstable D-branes
\cite{Kutasov:2003er,Niarchos:2004rw}
 it is  known that the effective action
of the type (\ref{act2}) is suitable for the
description of the tachyon dynamics close to
the marginal profile. This conclusion implies
that it is not quiet correct to perform standard
 analysis of fluctuations  when we 
split the general field $T$ into the
part obeying the classical equation of motion 
$T_c$ and the fluctuation part $t$  as  $T=T_c+t$
and we also
presume that  fluctuations are small
with respect to the classical part $T_c$. 
In order to study the
fluctuations around the classical
solution we will rather consider   field
$T$ in the form  
\begin{equation} \label{fluct}
T=t(x^{\mu})T_{c} \ , T_c=\lambda e^{\beta_{\mu}x^{\mu}} \ ,
\beta_{\mu}=V_{\mu} \ . 
\end{equation}
For such a field  we get
\begin{eqnarray}
\bb=1+e^{-2\Phi}\left(4T^2+\eta^{\mu\nu}
\partial_{\mu}T\partial_{\nu}T-2T\eta^{\mu\nu}
\partial_{\mu}TV_{\nu}\right)=\nonumber \\
=1+e^{-2\Phi}\left[t^2(4T_c^2+\eta^{\mu\nu}
\partial_{\mu}T_c\partial_{\nu}T_c-2
\eta^{\mu\nu}T_c\partial_{\mu}T_cV_{\nu})
\right.
\nonumber \\
\left.+T^2_c(\eta^{\mu\nu}\partial_{\mu}t
\partial_{\nu}t-2t\eta^{\mu\nu}\partial_{\nu}t
V_{\nu})+2tT_c\eta^{\mu\nu}\partial_{\mu}T_c
\partial_{\nu}t\right]=
\nonumber \\
=1+e^{-2\Phi}T^2_c[\eta^{\mu\nu}\partial_{\mu}t
\partial_{\nu}t+t\eta^{\mu\nu}\partial_{\mu}t
(\beta_{\nu}-V_{\nu})]=
1+\lambda^2\eta^{\mu\nu}\partial_{\mu}t\partial_{\nu}t 
\nonumber \\
\end{eqnarray}
After including the contribution from
the term (\ref{slambda}) that is equal
for the linear dilaton background to $-A$
we obtain  the action for fluctuation modes
around the classical solution $T_c$ in the form
\begin{equation}\label{actfluc}
S_t=-A\int d^2x \left(\sqrt{1+\lambda^2\eta^{\mu\nu}\partial_{\mu}t
\partial_{\nu}t}-1\right) \ .
\end{equation}
The equation of motion that arises from
(\ref{actfluc}) is
\begin{eqnarray}\label{fluceq}
\partial_{\mu}\left[\frac{\eta^{\mu\nu}\partial_{\nu}t}
{\sqrt{1+\lambda^2
\eta^{\mu\nu}\partial_{\mu}t\partial_{\nu}t}}\right]=0
\Rightarrow \nonumber \\
\Rightarrow
\frac{\partial_{\mu}\eta^{\mu\nu}\partial_{\nu}t}
{\sqrt{1+\lambda^2\eta^{\mu\nu}\partial_{\mu}t\partial_{\nu}t}}
-\lambda^2
\frac{\partial_{\mu}[\eta^{\kappa\rho}\partial_{\kappa}t
\partial_{\rho}t]\eta^{\mu\nu}\partial_{\nu}t}
{2(1+\lambda^2\eta^{\mu\nu}\partial_{\nu}t\partial_{\nu}t)^{3/2}}=0
\nonumber \\
\end{eqnarray}
For the plane-wave mode
 $t_k=e^{ik_{\mu}x^{\mu}}$ we obtain
from (\ref{fluceq}) 
\begin{eqnarray}
-\frac{t_kk_{\mu}k_{\nu}\eta^{\mu\nu}}{
\sqrt{1-\lambda^2\eta^{\mu\nu}k_{\mu\nu}t_k^2}}
+\lambda^2\frac{\partial_{\mu}[k_{\rho}k_{\kappa}\eta^{\kappa\rho}
t_k^2]\eta^{\mu\nu}ik_{\nu}t_k}
{2(1-\lambda^2
\eta^{\mu\nu}k_{\nu}k_{\nu}t_k^2)^{3/2}}=0
\Rightarrow \nonumber \\
-\frac{t_kk_{\mu}k_{\nu}\eta^{\mu\nu}}{
\sqrt{1-\lambda^2\eta^{\mu\nu}k_{\mu}k_{\nu}t_k^2}}
-\lambda^2\frac{k^2 k_{\mu}k_{\nu}\eta^{\mu\nu}t_k^3}
{(1-\lambda^2\eta^{\mu\nu}k_{\mu}k_{\nu}t_k^2)^{3/2}}=0
\Rightarrow k_{\mu}k_{\nu}\eta^{\mu\nu}=0 
\nonumber \\
\end{eqnarray}
so we get the condition that the fluctuation mode
is massless which agrees with standard
analysis of particle spectrum in two dimensional
string theory. 
 However we must stress one important point.
As is well known from the open string case 
\cite{Kutasov:2003er,Niarchos:2004rw} the tachyon
effective action of type (\ref{act3}) is presumed
to correctly describe the tachyon dynamics in the
vicinity of the marginal tachyon profile only.
Then one can argue that the fluctuations modes
$t$ should have small momenta $k_{\mu}\ll 1$ or
equivalently $\partial_{\mu} t\ll 1$. In order
to describe fluctuations modes with larger
derivatives 
one should consider  more general form of the 
tachyon effective action where  terms
with derivatives of higher order are included. 
\section{Conclusion}\label{fifth}
In this paper we have suggested the field
theory effective action for closed string tachyon that
has the marginal tachyon profile as its exact 
solution. Since generally the tachyon marginal
perturbation comes with connection with 
the nontrivial dilaton we have studied
this action in the linear dilaton background.
We have obtained this action by direct 
generalisation of the analysis performed in
\cite{Kluson:2004qy}.  Then we have mainly focused
on the two dimensional string theory. We have
seen that in two dimensions the analysis
simplifies considerably. Moreover, after inclusion
of the constant term into the tachyon effective action
we have got the action for metric, dilaton and tachyon
that has the linear dilaton background with spacelike
dilaton, flat spacetime  metric and with marginal tachyon
profile as its exact solution. We would
like to stress that this is the background 
that corresponds to the exact  
Liouville conformal field theory.  
We mean that this is very interesting property of this
model. We have also found
another exact solution of this combined system
however they meaning is
 unclear   since they contain time-dependent
dilaton and tachyon fields. According
to our remark given above this background should
corresponds to some form of time-like Liouville field
theory which is not completely understood at
present \cite{Strominger:2003fn,Schomerus:2003vv}.
 We have also found exact solution
of the tachyon effective action
that has the same form as the
tachyon profile
 that plays
significant role in the establishing the
 correspondence between
 spacetime two dimensional
string theory and its dual matrix model
 \cite{Polchinski:uq}.
 Unfortunately
we have found that this background  generates nonzero
dilaton source that diverges for $x\rightarrow \pm \infty$
and hence cannot be considered as an
 exact solution of our toy model of two
dimensional string theory. 

Let us mention some open problems that deserve
to be studied. It is clear that our model
can be used for description of type 0A and 
0B theories 
\cite{Douglas:2003up,Takayanagi:2003sm,
Klebanov:2003wg,Giveon:2003wn,Kapustin:2003hi,Gukov:2003yp}.
We mean that would be interesting
to study whether our model allows more general
solution with nontrivial  metric and dilaton together
with nonzero tachyon and find their possible
relations to the solutions of two dimensional
string theory that were found in the past
\footnote{For some recent papers, see 
\cite{Davis:2004xb,Thompson:2003fz,
Strominger:2003tm,Giveon:2003wn}, where
also extensive list of references can be found.}.
 It would be also interesting
to study this model in $D$-dimensional  space-times with
$D>2$ where the tachyon potential does not vanish. We
hope to return to these problems in future.
\\   
\\
{\bf Acknowledgement}

This work was supported by the
Czech Ministry of Education under Contract No.
14310006.
\\
\\


\begin{thebibliography}{20}
\bibitem{Sen:1999mg}
A.~Sen,
\emph{``Non-BPS states and branes in string theory,''}
arXiv:hep-th/9904207.

\bibitem{Nakayama:2004vk}
Y.~Nakayama,
\emph{``Liouville field theory: 
A decade after the revolution,''}
arXiv:hep-th/0402009.



\bibitem{Taylor:2002uv}
W.~Taylor,
\emph{``Lectures on D-branes, 
tachyon condensation, and string field theory,''}
arXiv:hep-th/0301094.

\bibitem{Arefeva:2001ps}
I.~Y.~Arefeva, D.~M.~Belov, A.~A.~Giryavets, A.~S.~Koshelev and P.~B.~Medvedev,
\emph{``Noncommutative field theories and 
(super)string field theories,''}
arXiv:hep-th/0111208.

\bibitem{Ohmori:2001am}
K.~Ohmori,
\emph{``A review on tachyon condensation 
in open string field theories,''}
arXiv:hep-th/0102085.



\bibitem{Schwarz:1999vu}
J.~H.~Schwarz,
\emph{``TASI lectures on non-BPS D-brane systems,''}
arXiv:hep-th/9908144.

\bibitem{Lerda:1999um}
A.~Lerda and R.~Russo,
\emph{``Stable non-BPS states in string theory: 
A pedagogical review,''}
Int.\ J.\ Mod.\ Phys.\ A {\bf 15} (2000) 771
[arXiv:hep-th/9905006].

\bibitem{Taylor:2003gn}
W.~Taylor and B.~Zwiebach,
\emph{``D-branes, tachyons, and string field theory,''}
arXiv:hep-th/0311017.




\bibitem{Adams:2001sv}
A.~Adams, J.~Polchinski and E.~Silverstein,
\emph{``Don't panic! Closed string tachyons in ALE space-times,''}
JHEP {\bf 0110}, 029 (2001)
[arXiv:hep-th/0108075].

\bibitem{Dabholkar:2001if}
A.~Dabholkar,
\emph{``Tachyon condensation and black hole entropy,''}
Phys.\ Rev.\ Lett.\  {\bf 88}, 091301 (2002)
[arXiv:hep-th/0111004].

\bibitem{Dabholkar:2001wn}
A.~Dabholkar and C.~Vafa,
\emph{``tt* geometry and closed string tachyon potential,''}
JHEP {\bf 0202}, 008 (2002)
[arXiv:hep-th/0111155].

\bibitem{Harvey:2001wm}
J.~A.~Harvey, D.~Kutasov, E.~J.~Martinec and G.~Moore,
\emph{``Localized tachyons and RG flows,''}
arXiv:hep-th/0111154.

\bibitem{Martinec:2002tz}
E.~J.~Martinec,
\emph{``Defects, decay, and dissipated states,''}
arXiv:hep-th/0210231.

\bibitem{Gutperle:2002ki}
M.~Gutperle, M.~Headrick, S.~Minwalla and V.~Schomerus,
\emph{``Space-time energy decreases under world-sheet RG flow,''}
JHEP {\bf 0301}, 073 (2003)
[arXiv:hep-th/0211063].

\bibitem{Armoni:2003va}
A.~Armoni, E.~Lopez and A.~M.~Uranga,
\emph{``Closed strings tachyons and non-commutative instabilities,''}
JHEP {\bf 0302}, 020 (2003)
[arXiv:hep-th/0301099].

\bibitem{He:2003yw}
Y.~H.~He,
\emph{ ``Closed string tachyons, 
non-supersymmetric orbifolds and generalised McKay
correspondence,''}
Adv.\ Theor.\ Math.\ Phys.\  {\bf 7}, 121 (2003)
[arXiv:hep-th/0301162].


\bibitem{DaCunha:2003fm}
B.~C.~Da Cunha and E.~J.~Martinec,
\emph{``Closed string tachyon 
condensation and worldsheet inflation,''}
Phys.\ Rev.\ D {\bf 68}, 063502 (2003)
[arXiv:hep-th/0303087].

\bibitem{Minwalla:2003hj}
S.~Minwalla and T.~Takayanagi,
\emph{``Evolution of D-branes under 
closed string tachyon condensation,''}
JHEP {\bf 0309}, 011 (2003)
[arXiv:hep-th/0307248].

\bibitem{Belopolsky:1994sk}
A.~Belopolsky and B.~Zwiebach,
\emph{ ``Off-shell closed string amplitudes: 
Towards a computation of the tachyon
potential,''}
Nucl.\ Phys.\ B {\bf 442}, 494 (1995)
[arXiv:hep-th/9409015].

\bibitem{Belopolsky:1994bj}
A.~Belopolsky,
\emph{``Effective Tachyonic potential in 
closed string field theory,''}
Nucl.\ Phys.\ B {\bf 448}, 245 (1995)
[arXiv:hep-th/9412106].

\bibitem{Okawa:2004rh}
Y.~Okawa and B.~Zwiebach,
\emph{``Twisted tachyon condensation 
in closed string field theory,''}
JHEP {\bf 0403} (2004) 056
[arXiv:hep-th/0403051].

\bibitem{Dabholkar:2004ky}
A.~Dabholkar, A.~Iqubal and J.~Raeymaekers,
\emph{``Off-shell interactions for 
closed-string tachyons,''}
arXiv:hep-th/0403238.

\bibitem{Kutasov:1991pv}
D.~Kutasov,
\emph{``Some properties of (non)critical strings,''}
arXiv:hep-th/9110041.

\bibitem{Martinec:1991kn}
E.~J.~Martinec,
\emph{``An Introduction to 2-d 
gravity and solvable string models,''}
arXiv:hep-th/9112019.

\bibitem{Klebanov:1991qa}
I.~R.~Klebanov,
\emph{``String theory in two-dimensions,''}
arXiv:hep-th/9108019.

\bibitem{Ginsparg:is}
P.~H.~Ginsparg and G.~W.~Moore,
\emph{``Lectures On 2-D Gravity 
And 2-D String Theory,''}
arXiv:hep-th/9304011.

\bibitem{DiFrancesco:1993nw}
P.~Di Francesco, P.~H.~Ginsparg and J.~Zinn-Justin,
\emph{``2-D Gravity and random matrices,''}
Phys.\ Rept.\  {\bf 254}, 1 (1995)
[arXiv:hep-th/9306153].

\bibitem{Jevicki:1993qn}
A.~Jevicki,
\emph{``Development in 2-d string theory,''}
arXiv:hep-th/9309115.

\bibitem{Polchinski:1994mb}
J.~Polchinski,
\emph{``What is string theory?,''}
arXiv:hep-th/9411028.



\bibitem{Karczmarek:2003pv}
J.~L.~Karczmarek and A.~Strominger,
\emph{``Matrix cosmology,''}
arXiv:hep-th/0309138.




\bibitem{Karczmarek:2004ph}
J.~L.~Karczmarek and A.~Strominger,
\emph{``Closed string tachyon condensation at c = 1,''}
arXiv:hep-th/0403169.

\bibitem{Das:2004hw}
S.~R.~Das, J.~L.~Davis, F.~Larsen and P.~Mukhopadhyay,
\emph{``Particle production in matrix cosmology,''}
arXiv:hep-th/0403275.



\bibitem{Strominger:2003fn}
A.~Strominger and T.~Takayanagi,
\emph{``Correlators in timelike bulk Liouville theory,''}
Adv.\ Theor.\ Math.\ Phys.\  {\bf 7} (2003) 369
[arXiv:hep-th/0303221].

\bibitem{Kluson:2003xn}
J.~Kluson,
\emph{``The Schrodinger wave 
functional and closed string rolling tachyon,''}
Int.\ J.\ Mod.\ Phys.\ A {\bf 19} (2004) 751
[arXiv:hep-th/0308023].

\bibitem{Schomerus:2003vv}
V.~Schomerus,
\emph{``Rolling tachyons from Liouville theory,''}
JHEP {\bf 0311} (2003) 043
[arXiv:hep-th/0306026].



\bibitem{Kutasov:2003er}
D.~Kutasov and V.~Niarchos,
\emph{``Tachyon effective actions in open string theory,''}
Nucl.\ Phys.\ B {\bf 666} (2003) 56
[arXiv:hep-th/0304045].



\bibitem{Sen:1999md}
A.~Sen,
\emph{``Supersymmetric world-volume 
action for non-BPS D-branes,''}
JHEP {\bf 9910} (1999) 008
[arXiv:hep-th/9909062].


\bibitem{Garousi:2000tr}
M.~R.~Garousi,
\emph{``Tachyon couplings on non-BPS 
D-branes and Dirac-Born-Infeld action,''}
Nucl.\ Phys.\ B {\bf 584} (2000) 284
[arXiv:hep-th/0003122].

\bibitem{Bergshoeff:2000dq}
E.~A.~Bergshoeff, M.~de Roo, T.~C.~de Wit, E.~Eyras and S.~Panda,
\emph{``T-duality and actions for non-BPS D-branes,''}
JHEP {\bf 0005} (2000) 009
[arXiv:hep-th/0003221].

\bibitem{Kluson:2000iy}
J.~Kluson,
\emph{``Proposal for non-BPS D-brane action,''}
Phys.\ Rev.\ D {\bf 62} (2000) 126003
[arXiv:hep-th/0004106].



\bibitem{Minahan:2000tg}
J.~A.~Minahan and B.~Zwiebach,
\emph{``Gauge fields and fermions 
in tachyon effective field theories,''}
JHEP {\bf 0102} (2001) 034
[arXiv:hep-th/0011226].


\bibitem{Sen:2002nu}
A.~Sen,
\emph{``Rolling tachyon,''}
JHEP {\bf 0204} (2002) 048
[arXiv:hep-th/0203211].

\bibitem{Sen:2002an}
A.~Sen,
\emph{``Field theory of tachyon matter,''}
Mod.\ Phys.\ Lett.\ A {\bf 17} (2002) 1797
[arXiv:hep-th/0204143].


\bibitem{Sen:2002qa}
A.~Sen,
\emph{``Time and tachyon,''}
Int.\ J.\ Mod.\ Phys.\ A {\bf 18} (2003) 4869
[arXiv:hep-th/0209122].



\bibitem{Sen:2003tm}
A.~Sen,
\emph{``Dirac-Born-Infeld 
action on the tachyon kink and vortex,''}
Phys.\ Rev.\ D {\bf 68} (2003) 066008
[arXiv:hep-th/0303057].

\bibitem{Smedback:2003ur}
M.~Smedback,
\emph{``On effective actions for 
the bosonic tachyon,''}
JHEP {\bf 0311} (2003) 067
[arXiv:hep-th/0310138].

\bibitem{Fotopoulos:2003yt}
A.~Fotopoulos and A.~A.~Tseytlin,
\emph{``On open superstring partition 
function in inhomogeneous rolling tachyon
background,''}
JHEP {\bf 0312} (2003) 025
[arXiv:hep-th/0310253].



\bibitem{Sen:2003zf}
A.~Sen,
\emph{``Moduli space of unstable D-branes on a 
circle of critical radius,''}
arXiv:hep-th/0312003.

\bibitem{Niarchos:2004rw}
V.~Niarchos,
\emph{``Notes on tachyon effective 
actions and Veneziano amplitudes,''}
arXiv:hep-th/0401066.

\bibitem{Lambert:2002hk}
N.~D.~Lambert and I.~Sachs,
\emph{``Tachyon dynamics and the effective 
action approximation,''}
Phys.\ Rev.\ D {\bf 67} (2003) 026005
[arXiv:hep-th/0208217].

\bibitem{Lambert:2001fa}
N.~D.~Lambert and I.~Sachs,
\emph{``On higher derivative terms 
in tachyon effective actions,''}
JHEP {\bf 0106} (2001) 060
[arXiv:hep-th/0104218].

\bibitem{Kluson:2004qy}
J.~Kluson,
\emph{``Proposal for the open string 
tachyon effective action in the linear dilaton
background,''}
arXiv:hep-th/0403124.



\bibitem{Kluson:2003sr}
J.~Kluson,
\emph{``Note on D-brane effective action 
in the linear dilaton background,''}
JHEP {\bf 0311} (2003) 068
[arXiv:hep-th/0310066].





\bibitem{Kluson:2004pj}
J.~Kluson,
\emph{``Bosonic D-brane effective 
action in linear dilaton background,''}
arXiv:hep-th/0401236.

\bibitem{Polchinski:uq}
J.~Polchinski,
\emph{``Classical Limit 
Of (1+1)-Dimensional String Theory,''}
Nucl.\ Phys.\ B {\bf 362} (1991) 125.




\bibitem{Douglas:2003up}
M.~R.~Douglas, I.~R.~Klebanov, 
D.~Kutasov, J.~Maldacena, E.~Martinec and N.~Seiberg,
\emph{``A new hat for the c = 1 matrix model,''}
arXiv:hep-th/0307195.

\bibitem{Takayanagi:2003sm}
T.~Takayanagi and N.~Toumbas,
\emph{``A matrix model dual 
of type 0B string theory in two dimensions,''}
JHEP {\bf 0307} (2003) 064
[arXiv:hep-th/0307083].

\bibitem{Klebanov:2003wg}
I.~R.~Klebanov, J.~Maldacena and N.~Seiberg,
\emph{``Unitary and complex 
matrix models as 1-d type 0 strings,''}
arXiv:hep-th/0309168.

\bibitem{Giveon:2003wn}
A.~Giveon, A.~Konechny, A.~Pakman and A.~Sever,
\emph{``Type 0 strings in a 2-d black hole,''}
arXiv:hep-th/0309056.

\bibitem{Kapustin:2003hi}
A.~Kapustin,
\emph{``Noncritical superstrings 
in a Ramond-Ramond background,''}
arXiv:hep-th/0308119.

\bibitem{Gukov:2003yp}
S.~Gukov, T.~Takayanagi and N.~Toumbas,
\emph{``Flux backgrounds in 2D string theory,''}
JHEP {\bf 0403} (2004) 017
[arXiv:hep-th/0312208].

\bibitem{Davis:2004xb}
J.~Davis, L.~A.~Pando Zayas and D.~Vaman,
\emph{``On black hole thermodynamics of 2-D type 0A,''}
JHEP {\bf 0403} (2004) 007
[arXiv:hep-th/0402152].

\bibitem{Thompson:2003fz}
D.~M.~Thompson,
\emph{``AdS solutions of 2D type 0A,''}
arXiv:hep-th/0312156.

\bibitem{Strominger:2003tm}
A.~Strominger,
\emph{``A matrix model for AdS(2),''}
JHEP {\bf 0403} (2004) 066
[arXiv:hep-th/0312194].


\end{thebibliography}
\end{document}